\documentstyle[psfig,12pt]{article}
\def\TL{\hfil$\displaystyle{##}$}
\def\TR{$\displaystyle{{}##}$\hfil}
\def\TC{\hfil$\displaystyle{##}$\hfil}
\def\TT{\hbox{##}}
\def\seqalign#1#2{\vcenter{\openup1\jot
  \halign{\strut #1\cr #2 \cr}}}

\def\fixit#1{}

\def\overleftrightarrow#1{\vbox{\ialign{##\crcr
     $\leftrightarrow$\crcr\noalign{\kern-0pt\nointerlineskip}
     $\hfil\displaystyle{#1}\hfil$\crcr}}}

\def\lsim{\mathrel{\mathstrut\smash{\ooalign{\raise2.5pt\hbox{$<$}\cr\lower2.5pt\hbox{$\sim$}}}}}
\def\gsim{\mathrel{\mathstrut\smash{\ooalign{\raise2.5pt\hbox{$>$}\cr\lower2.5pt\hbox{$\sim$}}}}}

\def\sqr#1#2{{\vcenter{\vbox{\hrule height.#2pt
         \hbox{\vrule width.#2pt height#1pt \kern#1pt
            \vrule width.#2pt}
         \hrule height.#2pt}}}}

\def\href#1#2{#2}  

\def\eqalign#1{\vcenter{\openup1\jot
    \halign{\strut\span\TL & \span\TR\cr #1 \cr
   }}}
\def\lbldef#1#2{\expandafter\gdef\csname #1\endcsname {#2}}
\def\eqn#1#2{\lbldef{#1}{(\ref{#1})}%
\begin{equation} \eqalign{#2} \label{#1} \end{equation}}
\def\eno#1{(\ref{#1})}

\def\eqn#1#2{\lbldef{#1}{(\ref{#1})}%
\begin{equation} \eqalign{#2} \label{#1} \end{equation}}
\def\frac#1#2{{#1 \over #2}}    
\begin{document}
\baselineskip=15.5pt
\pagestyle{plain}
\setcounter{page}{1}

\begin{titlepage}

\begin{flushright}
CALT-68-2351 \\
CITUSC/01-035 \\
hep-th/0110193
\end{flushright}
\vfil

\begin{center}
{\huge On non-uniform black branes}
\end{center}

\vfil
\begin{center}
{\large Steven S. Gubser$^{\,}$\footnote{E-mail: 
{\tt ssgubser@theory.caltech.edu}}$^,$\footnote{On leave from Princeton
University.}}
\end{center}

$$\seqalign{\span\TL & \span\TT}{
& Lauritsen Laboratory of Physics, 452-48 Caltech, Pasadena, CA  91125
}$$
\vfil

\begin{center}
{\large Abstract}
\end{center}

\noindent 
 Certain black branes are unstable toward fluctuations that lead to
non-uniform mass distributions.  We study static, non-uniform
solutions that differ only perturbatively from uniform ones.  For
uncharged black strings in five dimensions, we find evidence of a
first order transition from uniform to non-uniform solutions.

\vfil
\begin{flushleft}
October 2001
\end{flushleft}
\end{titlepage}
\newpage
\renewcommand{\thefootnote}{\arabic{footnote}}
\setcounter{footnote}{0}
\tableofcontents
\section{Introduction}
\label{Introduction}

The Gregory-Laflamme instability \cite{glOne,glTwo} has been explored
principally via linearized perturbation theory and thermodynamic
scaling arguments.  In its simplest incarnation, it is a tachyonic
mode in perturbations around an uncharged black brane which makes the
horizon bulge out at some points and squeeze in at others.  It is a
purely classical effect in Minkowskian signature, and we may think of
it crudely as a black brane redistributing its mass in order to gain
entropy.  Until recently it was commonly supposed that this evolution
would proceed until the horizon bifurcated into an array of black
holes (and then these black holes might run into one another,
clustering mass into ever fewer, ever larger black holes).

This naive view has recently been called into question because of a
result of Horowitz and Maeda \cite{hm}.  These authors demonstrated
rigorously that a black hole horizon {\it cannot} pinch off in finite
affine time, and they give less airtight arguments that such a
pinch-off also cannot happen in infinite affine time.  Thus it becomes
plausible that the endpoint of the evolution is a stable, non-uniform
black brane.\footnote{Something not too distant from this picture was
suggested in section~3 of \cite{cg}, where it was also suggested that
non-uniformity would develop through a first order transition.  The
reasoning there was rather different, and much less rigorous.}  In the
case of a black string, the non-uniform solution might be expected to
look like an unbroken string of identical beads.

Briefly, the aim of this paper is to look for such a non-uniform black
string using perturbation theory around the uniform solution, and to
discuss possible phase transitions between different types of
solutions.

The organization of the paper is as follows: in section~\ref{Order},
we discuss order parameters.  In section~\ref{Numerics}, we perform an
explicit perturbation analysis around the black string solution in
five dimensions.  Non-uniform solutions are constructed perturbatively
around the uniform string with $\eta = \eta_c$.  Unfortunately, even
this involves some numerical work, which will be explained in some
detail in sections~\ref{Olambda}-\ref{Improve}.  In
section~\ref{Discuss} we make some general remarks about how the
Landau theory of phase transitions might bear on non-uniform branes
of various dimensionalities, and also comment on possible
instabilities which could arise {\it after} a brane has become
non-uniform.  We summarize our key results on the five-dimensional
uncharged black string in section~\ref{Conclude}.

\section{Order parameters}
\label{Order}

The Gregory-Laflamme instability is an infrared effect: when it is
present, {\it all} accessible wavelengths below a certain threshold
are unstable.  Life is simpler if one imposes an infrared regulator:
for example, one may compactify the directions parallel to the black
brane on some torus.  Then the instability shows up only if the
Schwarzschild radius is sufficiently small compared with the size of
the torus (the zero-mode cannot be excited because of energy
conservation).  The commonly supposed endpoint of the evolution
(before \cite{hm}) was a single black hole, localized in the torus
directions as well as the orthogonal directions.  By a simple scaling
argument, such a black hole has more entropy for a given mass than the
uniform black brane starting point, provided the size of the torus is
sufficiently large compared to the Schwarzschild radius of the black
brane.

As long as we are discussing solutions of the classical empty space
Einstein equations, $R_{\mu\nu} = 0$, there is no intrinsic scale in
the problem.  This may be seen from the fact that a rigid rescaling of
the entire metric takes solutions into solutions.  The uniform black
brane determines a length scale, namely its Schwarzschild radius,
$R_{\rm Schwarzschild}$.  After compactifying the uniform black brane
on a torus, suppose the longest wavelength perturbation accessible has
wave number $k_{\rm min}$.  Then we may define the figure of merit
  \eqn{EtaDef}{
   \eta = R_{\rm Schwarzschild} k_{\rm min} \,.
  }
 For a non-uniform solution, we may still define $\eta$ as the value
that would pertain if the non-uniform brane were replaced by a uniform
one of the same mass.

If the black brane does become non-uniform, there is another
dimensionless number, $\lambda$, which describes how non-uniform it
is.  Supposing that the rotational symmetry in the directions
transverse to the brane is preserved, we can define unambiguously the
area of a slice of the horizon over any point along the brane, and
hence an effective Schwarzschild radius at any point.  Then we can
define
  \eqn{LambdaDef}{
   \lambda = {1 \over 2} \left( {R_{\rm Schwarzschild,\ max} \over
    R_{\rm Schwarzschild,\ min}} - 1 \right) \,.
  }
 Thus $\lambda=0$ for the uniform black brane; $\lambda = \infty$ for
a black hole localized on the torus, and $\lambda$ is finite for a
non-uniform black brane with an unbroken horizon.

A final meaningful number for a non-uniform horizon compactified on a
torus (or other manifold for that matter) is the number $n$ of local
maxima of the Schwarzschild radius---again on the assumption that
transverse rotational symmetry is preserved, so that the Schwarzschild
radius has a good local definition everywhere on the brane.

Roughly speaking, the problem gets harder as $\eta$ decreases.  For
$\eta$ large, the compactification scale is much smaller than another
scale in the problem, so we may expect the uniform black brane to be
stable and to win out over any other solution.  Below some critical
$\eta_c$, the uniform black brane becomes unstable to small
perturbations: this is the Gregory-Laflamme instability.  What happens
for smaller $\eta$ is not known.  Focusing first on the black string,
one might reasonably expect that for some range of $\eta$ below
$\eta_c$, there is a stable solution with finite $\lambda$ and $n=1$.
Then the question becomes, does non-zero $\lambda$ develop smoothly or
suddenly as one crosses $\eta_c$?  That is, is the phase transition
from uniform to non-uniform black strings continuous or first
order?\footnote{It makes sense to speak of phase transitions even
though the system is of finite extent because it is much larger than
Planck scale.}  The results of section~\ref{Numerics} will suggest
that it is first order.

\section{Numerics on the black string}
\label{Numerics}

\subsection{Generalities}

At $\eta=\eta_c$, the perturbative Gregory-Laflamme instability
becomes a zero-mode of the uniform black brane solution, so we may
expect that there is a new branch of the solution space heading out to
nonzero $\lambda$, but emanating from $\lambda=0$ and $\eta=\eta_c$.
This new branch has one maximum for the Schwarzschild radius ($n=1$)
because the zero mode changes $R_{\rm Schwarzschild}$,
multiplicatively, by an amount $1 + \lambda \cos kx$ where $k$ is the
fundamental wave-number of the black string.\footnote{We defined the
$1/2$ in \LambdaDef\ so that it would coincide with the usage of
$\lambda$ in this section, {\it to lowest order in small $\lambda$.}
We will not be concerned with a more precise translation of
definitions.}  A continuous moduli space may seem at odds with first
order behavior; see however figure~\ref{figB}(a),
equation~\eno{EtaSigmaExpand}, and the surrounding discussion for a
preview of how the two are consistent.

We want static five-dimensional black string solutions to Einstein's
equations in the absence of matter: $R_{\mu\nu} = 0$.  A general
ansatz for such solutions is
  \eqn{BSAnsatzFirst}{
   ds^2 = -e^{2A} \left( 1 - {2M \over r} \right) dt^2 + 
    e^{2B} \left[ {dr^2 \over 1 - {2M \over r}} + dz^2 \right] + 
    e^{2C} r^2 d\Omega_2^2 \,,
  }
 where $A$, $B$, and $C$ are arbitrary functions of $r$ and $z$.  Here
we have used diffeomorphism freedom to constrain the ratio of $g_{rr}$
and $g_{zz}$, to get rid of any $g_{rz}$ term, and to ensure that the
black string horizon is at $r = 2M$.  The ansatz \BSAnsatzFirst\ is
almost in conformal gauge: it would be exactly in conformal gauge if
we changed coordinates to $\tilde{r}$ such that $d\tilde{r}^2 = dr^2 /
(1 - 2M/r)$.\footnote{We are able to constrain the horizon to lie at a
fixed value of $\tilde{r}$ {\it after} fixing the conformal gauge
because this gauge is preserved by conformal transformations.  An
appropriate conformal map will indeed send the region outside the
horizon to the half-plane $\tilde{r} > 0$, or $r > 2M$.}  Clearly, if
$A=B=C=0$, we have the uniform black string.

It's possible to change coordinates and rescale the entire metric so
that it becomes
  \eqn{BSAnsatz}{
   ds^2 = -e^{2A} \left( 1 - {1 \over y} \right) dt^2 + 
    e^{2B} \left[ {dy^2 \over 1 - {1 \over y}} + dx^2 \right] + 
    e^{2C} y^2 d\Omega_2^2 \,.
  }
 The Schwarzschild radius of the uniform solution is $1$.  The
periodicity of the $x$ direction is $2\pi/\eta$, so $\eta$ coincides
with the minimal wave-number $k$ which we can excite.  Looking for
solutions of the form \BSAnsatz\ to $R_{\mu\nu} = 0$ amounts to an
elliptic boundary value problem: boundary conditions are imposed at
the horizon to ensure that it is regular, and other boundary
conditions are imposed at infinity to ensure that the solution is
asymptotically flat.  (Actually, we will see that regularity plus
asymptotic flatness are not quite sufficient to fix all integration
constants.  This subtlety will figure prominently in
sections~\ref{OlambdaSquared} and~\ref{HigherOrder}, but may otherwise
be glossed over.)  Elliptic boundary value problems for {\it ordinary}
differential are fairly straightforward to deal with: the basic
strategy is some form of ``shooting'' algorithm, where one
artificially imposes full Cauchy data at one end of the integration
region, then integrates the ODE's to the other end, and then checks to
see whether the boundary conditions are satisfied there.  To
understand why elliptic boundary problems are harder for {\it partial}
differential equations, imagine trying the same strategy, for example
imposing Cauchy data at $y=1$ and then integrating out to infinity.
One has to discretize the $x$-interval to perform this integration.
Intuitively, such a discretization involves about as many Fourier
modes as there are pixels.  The higher Fourier modes have higher mass.
Each of these modes has a solution which grows exponentially at
infinity rather than vanishing.  Round-off error will tend to source
these growing solutions, with the result that numerical integration
outward in $y$ quickly diverges, for any initial data at the horizon.
It appears in fact that there is no standard algorithm for solving
elliptic boundary value problems for non-linear PDE's.  A custom
algorithm is usually constructed, based commonly on some global
relaxation method, to deal with a specific problem.

Instead of pursuing the intensive numerical route, let us make an
expansion around the uniform string:
  \eqn{ABCexpand}{
   A &= \lambda^2 a_0(y) + \lambda a_1(y) \cos kx + 
        \lambda^2 a_2(y) \cos 2kx + O(\lambda^3)  \cr
   B &= \lambda^2 b_0(y) + \lambda b_1(y) \cos kx + 
        \lambda^2 b_2(y) \cos 2kx + O(\lambda^3)  \cr
   C &= \lambda^2 c_0(y) + \lambda c_1(y) \cos kx + 
        \lambda^2 c_2(y) \cos 2kx + O(\lambda^3) \,.
  }
 We will work at small $\lambda$, but the coefficient functions are
finite.  The expansion \ABCexpand\ is appropriate for studying
perturbations at the wavelength which is marginally stable.  The
reason is that at lowest order, $O(\lambda)$, a perturbation at this
wavelength is a zero mode of the uniform black brane.  At
$O(\lambda^2)$, back-reaction leads to non-trivial $X_0$ and $X_2$,
where $X = a$, $b$, or $c$.  If we make a Kaluza-Klein reduction in
the $x$ direction, then the $X_i$ for $i \neq 0$ are massive modes,
while the $X_0$ are massless modes.  Thus the $X_i$ for $i \neq 0$
fall off exponentially for large $y$, while the $X_0$ fall off as
inverse powers of $y$.  So to read off the mass density of the black
string, all we need is the $x$-independent modes.  We will elaborate
on this point later.

Having made the expansion \ABCexpand, we are left with a finite set of
ODE's at each order in $\lambda$.  Boundary conditions still must be
imposed both at the horizon and at infinity.  Let us now enter into
the details of the calculation.\footnote{Readers interested in a
fuller description can examine the Mathematica notebook
\cite{NBposted}.}  The reader uninterested in technical points may
wish to skip directly to section~\ref{Interpret}.

\subsection{Solution to $O(\lambda)$}
\label{Olambda}

The basic ODE's can be derived by plugging \ABCexpand\ into the
Einstein equations, $R_{\mu\nu} = 0$.  At $O(\lambda)$, one can solve
algebraically for $b_1(y)$, with the result
  \def\frac#1#2{{#1 \over #2}}
  \eqn{bZeroSolve}{
   b_1 = -\frac{a_1 + 2\,\left( -1 + y \right) \,
       \left( 2\,c_1 + y\,a_1' + 2\,y\,c_1' \right) }
    {3 - 4\,y} \,,
  }
 where primes denote $d/dy$.  This equation incorporates the zero
energy constraint.  One may also show that the $O(\lambda)$ equations
imply
  \eqn{bpZeroSolve}{
   b_1' = \frac{2}{-3 + 4\,y} \left( k^2\,y^2\,a_1 + 2\,b_1 - 2\,c_1 + 
       2\,k^2\,y^2\,c_1 + 2\,a_1 - 2\,y\,a_1 + 
       c_1 - 2\,y\,c_1 \right) \,,
  }
 which is a relation which we will find helpful in
section~\ref{OlambdaSquared}.  Eliminating $b_1$ from the other
differential equations at $O(\lambda)$ leads to
  \eqn{acOneODEs}{
   2\,k^2\,y^2\,a_1 + \left( 1 - 4\,y \right) \,a_1' - 
   2\,\left( c_1' + \left( -1 + y \right) \,y\,a_1'' \right) &= 0  \cr
  -2\,a_1 + \left( 2 + k^2\,\left( 3 - 4\,y \right) \,y^2 \right) \,c_1 + 
   3\,a_1' - 3\,y\,a_1' \qquad & \cr + 9\,c_1' - 16\,y\,c_1' + 
   8\,y^2\,c_1' - y (-1+y) (3 - 4\,y) \,c_1'' &= 0
  }
 Evidently, the horizon is a regular singular point of these
differential equations, which are linear because at $O(\lambda)$ we're
dealing precisely with linearized perturbations around the uniform
black string.  We can exploit the linearity to get rid of one
arbitrary constant: let us say $c_1(1) = 1$.  Examining the region
near $y=1$, one can show that two solutions of \acOneODEs\ are
singular.  Regularity at the horizon demands that these singular
solutions should be absent.  To specify Cauchy data at the horizon we
must therefore specify just one more quantity at the
horizon---conveniently $a_1(1)$.  Also $k$ must be specified.  At
infinity there are two exponentially growing solutions to \acOneODEs.
The natural boundary conditions at infinity are that these solutions
must be absent.  Thus we have two free parameters, $a_1(1)$ and $k$,
and two boundary conditions at infinity to satisfy.  An appropriate
two-parameter shooting algorithm is easy to implement.  The only
subtlety worth mentioning is that because we have {\it two} growing
solutions at infinity, it's possible to them to compete so that
$a_1(y)^2 + c_1(y)^2$ is small at a single large value of $y$, but
large at other values.  So to test whether one has eliminated both
growing solutions, an efficient and robust method is to integrate
$a_1(y)^2 + c_1(y)^2$ over some interval near the end of the
integration region.  We have done this, with the result $a_1(1) =
-0.552$ and $k=0.876$.  This value of $k$ is in accord with the
results of \cite{glOne}.\footnote{In comparing with Figure~1 of
\cite{glOne}, one must note that $D=4$ in \cite{glOne} corresponds to
the five-dimensional black string in our language.  Also, it appears
that the numerical integrations in \cite{glOne} were performed at
Schwarzschild radius $r_+ = 2$ (see also the remarks on p.~17 of
\cite{glTwo}).  To convert to $r_+ = 1$, $\Omega$ and $\mu$ should be
doubled.  And $\mu$ is what we have called $k$.  Thus the intersection
of the interpolating curve with the horizontal axis at $\mu \approx
0.43$ in their Figure~1 is indeed in accord with $k =
0.876$.}

\subsection{Mass, entropy, and temperature}

Before going on to $O(\lambda^2)$, let us describe in more detail how
the mass and entropy can be calculated.  The results in this section
are valid at any order in $\lambda$, and indeed make almost no
reference to the expansion \ABCexpand.  Following \cite{hh}, we define
the mass as
  \eqn{EDef}{
   E = -{1 \over 8\pi} \int_\Sigma \left( {}^3 K - {}^3 K_0 \right) \,,
  }
 where ${}^3 K$ is the trace of the extrinsic curvature of a surface
$\Sigma$ at constant $y$ and $t$, computed with respect to the
four-dimensional spatial metric: that is,
  \eqn{KDef}{
   {}^3 K = {1 \over \sqrt{{}^4 g}} \partial_i \left( \sqrt{{}^4 g} N^i 
    \right) \,,
  }
 where ${}^4 g$ is the metric on a $t = const$ spatial slice, and
$N^i$ is the unit normal to the surface $y = const$ in this slice.
The quantity ${}^3 K_0$ is the extrinsic curvature for a reference
geometry, in our case flat space.  A large $y$ limit is taken after
subtracting the contribution of the reference geometry, and then the
energy should be finite.  In making the subtraction, the surface
$\Sigma$ must be chosen to have the same induced geometry in the black
string background and the reference geometry, at least up to a
sufficiently high order in large $y$.  In our case, we may altogether
ignore the functions $X_i$ for $i\neq 0$ (effectively setting them
equal to $0$) because at large $y$ they fall off exponentially in $y$,
whereas the terms in ${}^3 K - {}^3 K_0$ that make finite
contributions to the energy are $O(1/y^2)$.  Let us {\it assume} the
leading order behavior
  \eqn{LeadingBC}{
   B = {B_\infty \over y} + O(1/y^2) \qquad 
    C = {C_\infty \log y \over y} + O(1/y) \,.
  }
 Then, suppressing all exponentially small corrections, we can write
  \eqn{KCompute}{\eqalign{
   \sqrt{{}^4 g} &= e^{2B+2C} {y^2 \over \sqrt{1 - 1/y}} \qquad
    N^y = e^{-B} \sqrt{1 - 1/y}  \cr
   {}^3 K &= {1 \over \sqrt{{}^4 g}} \partial_y 
     \left( \sqrt{{}^4 g} N^y \right)
     = {\sqrt{1 - 1/y} \over y^2} e^{-2B-2C} \partial_y 
        \left( e^{B+2C} y^2 \right)  \cr
     &= {2 \over y} - 2 C_\infty {\log y \over y^2} -
         {1 \over y^2} (1 + 3B_\infty - 2C_\infty) + 
         O(1/y^3) \,,
  }}
 This is to be compared with ${}^3 K_0 = 2/y_0$ for the flat reference
spacetime.  However, in making the subtraction one does not want $y_0
= y$, but rather $y_0 = y e^{C} = y + C_\infty \log y + O(1/y)$ so
that the $S^2$ has the same curvature in the black string geometry and
the reference geometry.  Thus we obtain
  \eqn{KZeroCompute}{\eqalign{
   {}^3 K_0 &= {2 \over y} - 2 C_\infty {\log y \over y^2}  \cr
   {}^3 K - {}^3 K_0 &= {1 \over y^2} (1 + 3B_\infty - 2C_\infty) + 
     O(1/y^3) \,.
  }}
 The area of the $S^2$ is $4\pi y^2$, and the length of the $S^1$ is
$2\pi/k$, so we see that the energy of the black string is $(1 +
3B_\infty - 2 C_\infty)/k$ up to overall constants.  In later sections
we will allow $k$ to vary, so it is more convenient to define $e =
k E/2\pi$ as the {\it average energy density}.  Then
  \eqn{deltaE}{
   {\delta e \over e} = 3 B_\infty - 2 C_\infty \,.
  }
 In applying the result \deltaE, it must be recalled that we {\it
assumed} \LeadingBC.  In a perturbative treatment in $\lambda$, one
must check this asymptotic behavior at any given order.  Most
modifications of the asymptotics would render the mass infinite.

Computing the entropy is more straightforward: it is just the horizon
area over $4 G_5$, so the fractional change in the average entropy
density, $s = k S/2\pi$, is
  \eqn{deltaS}{
   {\delta s \over s} = \left\langle e^{B+2C} \Big|_{y=1} 
     \right\rangle - 1 \,,
  }
 where $\langle\rangle$ means to average over the $x$ direction.  Thus
$(\delta s)/s$ winds up being some combination of the $X_i$ evaluated
at $y=1$.

The temperature is also easy to compute, using for instance the
standard prescription of rotating to Euclidean signature and demanding
no conical deficit at the horizon.  The result is
  \eqn{TDef}{
   {\delta T \over T} = e^{A-B} \Big|_{y=1} - 1
  }
 The right hand side is not manifestly independent of $x$.  A
convenient check on boundary conditions at the horizon is to verify
that they preclude any contribution to $(A-B) \big|_{y=1}$ other than
from zero modes.

\subsection{The zero modes at $O(\lambda^2)$}
\label{OlambdaSquared}

As per the discussion of the previous section, to compute $(\delta
e)/e$ and $(\delta s)/s$ to $O(\lambda^2)$, all we require is the
functions $X_0$ to this order (where as usual, $X$ can be $a$, $b$, or
$c$).  So we postpone discussion of the $X_2$ until
section~\ref{HigherOrder}.

The equations for the zero modes at $O(\lambda^2)$ are rather
complicated.  All the relevant equations take the schematic form $X_0
= X_1^2$, where the left hand side denotes some linear combination of
$a_0$, $b_0$, $c_0$ and their first and second derivatives, and the
right hand denotes a quadratic expression in $a_1$, $b_1$, $c_1$ and
their first and second derivatives.  Using \bZeroSolve, \bpZeroSolve,
and \acOneODEs, one can eliminate all second derivatives from the
right hand side of $X_0 = X_1^2$, and also eliminate all dependence on
$b_1$ and $b_1'$.  (This is a good idea for numerics, because the
functions $X_1$ are known only numerically, and differentiating them
leads to more numerical noise.)  One of the resulting equations is
  \eqn{bZeroODE}{\eqalign{
   -{d \over dy} 2 (-1+y) y {d \over dy} b_0 &= 
   \frac{2\,y}{{\left( 3 - 4\,y \right) }^2}
\Big( 2\,k^2\,y\,\left( 3 - 7\,y + 4\,y^2 \right) \,{a_1}^2 + 
      k^2\,y\,\left( -3 + 4\,y \right) \,{c_1}^2  \cr & \quad{} - 
      2\,\left( -1 + y \right) \,c_1\,\left( a_1' + 2\,c_1' \right)
       \cr & \quad{} - 
      \left( 3 - 5\,y + 2\,y^2 \right) \,\left( 2\,\left( -1 + y \right) \,a_1' - 
         c_1' \right) \,\left( a_1' + 2\,c_1' \right)
       \cr & \quad{} + 
      a_1\,\Big( k^2\,y\,\left( -3 + 10\,y - 8\,y^2 \right) \,c_1 + 
         2\,\left( -1 + y \right) \,\left( a_1' + 2\,c_1' \right)  \Big)  \Big) .
  }}
 This equation is easy to integrate twice numerically.  In order to
have regularity at the horizon, $b_0(1)$ must be finite, and this
fixes a constant in the first integration.  The constant in the second
integration can be fixed by demanding $b_0 \to 0$ as $y \to \infty$.
This boundary condition does not come from asymptotic flatness, but it
could be arranged through some rescalings.  No shooting is necessary.
After integrating, one can easily extract $B_\infty \approx 0.41
\lambda^2$.

The other equations derivable at $O(\lambda^2)$ for the zero modes
take the form
  \eqn{acZeroODEs}{\eqalign{
   a_0' - 4\,y\,a_0' - 2\,c_0' + 2\,y\,a_0'' - 
  2\,y^2\,a_0'' &= S_1(a_1,c_1)  \cr
   -4\,c_0 + 2\,a_0' - 2\,y\,a_0' + 6\,c_0' - 
  8\,y\,c_0' + 2\,y\,c_0'' - 2\,y^2\,c_0'' &=
     S_2(a_1,c_1;b_0)  \cr
   8\,c_0 - 8\,a_0' + 8\,y\,a_0' - 4\,c_0' + 
  8\,y\,c_0' &= S_3(a_1,c_1;b_0) \,,
  }}
 where the source terms $S_1$, $S_2$, and $S_3$ are quadratic
expressions in $a_1$, $c_1$ and their first derivatives, like the
right hand side of \bZeroODE.  $S_2$ and $S_3$ also have {\it linear}
dependence on $b_0$.  We regard $b_0$ as fixed from the considerations
of the previous paragraph.  Note that $a_0$ appears only through its
derivatives.  An additive shift of $a_0$ corresponds to a
multiplicative redefinition of $t$.  The equations in \acZeroODEs\ are
not independent: the first, for example, follows from the second and
the third together.  For numerics, it would seem most straightforward
to keep the third equation and drop either the first or the second,
since the third is first order.  However, it turns out to be better to
drop the third equation and integrate the first two
numerically.\footnote{T.~Wiseman pointed out to me that this
alternative led to considerably more tractable numerics near the
horizon.}  In order to do this we need Cauchy data at the horizon.  If
we assume regularity of $a_0$, $c_0$ and their first and second
derivatives, then by setting $y=1$ in the first and third equations in
\acZeroODEs\ we obtain
 \eqn{zBCs}{
  -3a_0'-2c_0' = S_1(a_1,c_1) \,, \qquad
    -4 c_0(1) - 2 c_0'(1) = S_2(a_1,c_1;b_0) \,,
 }
where all quantities are evaluated at $y=1$.  The third equation in
\acZeroODEs\ does not lead to any additional boundary conditions at
the horizon.  Two more boundary conditions are required to complete
the Cauchy data: one of them is the trivial additive constant on
$a_0$, which can be adjusted {\it a posteriori} so that $a_0(y) \to 0$
as $y \to \infty$; and the other is the value of $c_0(1)$.  Any choice
of $c_0(1)$ leads to a solution that is regular everywhere, with
asymptotics as in \LeadingBC.  The $O(\lambda^2)$ contribution to
$C_\infty$ depends on $c_0(1)$.

Clearly this leaves us with a puzzle: why is $C_\infty$, and hence the
mass, indeterminate?  Roughly, the answer is that, at the order to
which we are working, we are still free to ``superpose'' an arbitrary
$O(\lambda^2)$ change in the mass of the black hole on top of the
change that the non-uniformity induces.  The value of $c_0(1)$ is
undetermined at $O(\lambda^2)$, and will be fixed, if at all, by
considerations at higher order.  (We will revisit this issue in
section~\ref{HigherOrder}.)  The equations \acZeroODEs\ with the $S_i$
set to $0$ correspond precisely to perturbations of a Schwarzschild
black hole in four dimensions in a gauge where the radial part of the
metric is exactly $dy^2/(1-1/y)$.  With this choice of gauge, an
increase of the mass of the black hole is expressed as a solution to
the homogeneous equations derived from \acZeroODEs\ with $c_0(y) \sim
C_\infty (\log y)/y$ for large $y$.  Any multiple of this solution to
the homogeneous equations can be added to a solution to the
inhomogeneous equations: that is the origin of the ambiguity in the
mass.

To proceed, let us set $c_0(1)=0$.  Numerical integration of
\acZeroODEs\ is now straightforward.  Summarizing the results of the
numerics so far:
  \eqn{NumericsSummary}{\eqalign{
   a_1(1) = -0.55 \qquad
   b_1(1) = -0.55  \qquad
   c_1(1) = 1  \cr
   a_0(1) = 0.53 \qquad
   b_0(1) = 0.77 \qquad
   c_0(1) = 0  \cr
   B_\infty = 0.41 \lambda^2 \qquad
   C_\infty = 0.29 \lambda^2 \,.
  }}
 Given these values, together with \deltaE\ and \deltaS, one may
evaluate, to $O(\lambda^2)$,
  \eqn{DeltaEDeltaS}{\eqalign{
   {\delta e \over e} &= 3 B_\infty - 2 C_\infty \approx 0.64 \lambda^2 
    \cr
   {\delta s \over s} &= \left[ b_0 + 2 c_0 + {1 \over 4} b_1^2 + 
     b_1 c_1 + c_1^2 \right]_{y=1} \lambda^2 \approx 
    1.29 \lambda^2 \,.
  }}
 Adjusting $c_0(1)$ changes $(\delta s)/s$ by exactly twice the amount
that it does $(\delta e)/e$: this is because the entropy-mass relation
for the four-dimensional Schwarzschild solution is $S \propto M^2$.
A quantity which is supposed to be independent of $c_0(1)$ is
  \eqn{DeltaES}{
   \Delta_{es} = 
    {\delta s \over s} - 2 {\delta e \over e} = \sigma_1 \lambda^2 
     \qquad\hbox{where}\quad \sigma_1 \approx 0.002 \,.
  }
 The coefficient $\sigma_1$ is remarkably small compared to the values
in \NumericsSummary.  This may lead us to suspect that the true value
is zero, so that the non-zero result in \DeltaES\ is pure round-off
error.  We will presently give an entirely analytical demonstration
that $\sigma_1=0$; however, let us for the moment examine how we would
judge just from the numerics whether $\DeltaES$ is consistent with a
zero result.  One way way to test this hypothesis is to vary $c_0(1)$
and see how much $\sigma_1$ changes.  Since all the values in
\NumericsSummary\ are on the order of unity, it would seem natural for
$c_0(1)$ to have a similar magnitude.  So the deviations in the result
that come from assigning $c_0(1)$ the values $\pm 1$ rather than $0$
are something like a standard deviation.  This ``standard deviation''
turns out to be roughly $0.02$, so our result is indeed consistent
with $\sigma_1=0$ (in fact, one might say that we were lucky to get a
value as small as the one in \DeltaES).  Another test is to upgrade
the numerics in various ways and see what changes.  The results quoted
in \NumericsSummary-\DeltaES\ come from the last of five iterations of
improvement, which added about one decimal place of accuracy.  The
first three iterations were about one standard deviation positive.

The bottom line is that numerics are consistent with the hypothesis
that, at $O(\lambda^2)$, a uniform black string and a non-uniform one
with the same mass also have the same entropy.  More precisely, the
non-uniform black string's entropy exceeds that of the uniform black
string only by about one standard deviation, as defined heuristically
in the previous paragraph.  Now let us show analytically that
$\Delta_{es} = o(\lambda^2)$, using the First Law of thermodynamics
(which certainly should hold for the non-uniform solutions at hand,
since as static solutions to the vacuum Einstein's equation they have
well-defined event horizons).  We will need to assume that, if the
asymptotic size of the $S^1$ is held fixed, then the mass is specified
unambiguously once one has specified $\lambda$.  That this is true
will emerge from the discussion in section~\ref{HigherOrder}.  Thus
there is a one-parameter curve of non-uniform solutions in the
$\eta$-$\lambda$ plane, emanating from the point $(\eta_c,0) \approx
(0.876,0)$.  Suppose we slide incrementally along this curve, away
from $(\eta_c,0)$.  Then the First Law says $dE = T_c dS$, where $T_c$
is the temperature at $(\eta_c,0)$.  At this point, the black hole is
uniform, so we may use the standard four-dimensional Schwarzschild
relation $T_c = E_c / (2S_c)$.  Rearranging the first law slightly, we
obtain $2 (dE)/E_c = (dS)/S_c$.  Thus indeed $\Delta_{es} = 0$, at
least to some lowest order.  Now we have to explain why this works to
$O(\lambda^2)$.  The reason is that the temperature deviates from
$T_c$ only at $O(\lambda^2)$, and then this small deviation is further
suppressed (to $O(\lambda^4)$, in fact) when one integrates the
infinitesimal form of the First Law out from $0$ to $\lambda$.

The assumption that the asymptotic size of the $S^1$ is constant is
needed in the above derivation because otherwise the First Law would
have an additional term corresponding to variations in this quantity.
The assumption that only a one-parameter family of non-uniform
solutions exists with fixed asymptotic size of $S^1$ is needed for the
same reason: otherwise there would effectively be another
thermodynamic observable that would contribute to the First Law.

Obviously, although we may feel relieved that our numerics through
$O(\lambda^2)$ passes one non-trivial analytical check, we should also
feel disappointed that no really meaningful numbers came out of the
analysis so far, beyond what is in the literature.  To do better we
must proceed at least to $O(\lambda^3)$.  There are some conceptual
subtleties involved, which we will discuss in the next section.

\subsection{Higher orders in perturbation theory}
\label{HigherOrder}

To organize higher order in perturbation theory, it will be useful to
consider the expansions
  \eqn{ABCagain}{\eqalign{
   A &= \sum_{n=0}^\infty \lambda^n A_n \cos nKx  \cr
   B &= \sum_{n=0}^\infty \lambda^n B_n \cos nKx  \cr
   C &= \sum_{n=0}^\infty \lambda^n C_n \cos nKx \,,
  }}
 where for $X=A$, $B$, and $C$ we have the additional expansions
  \eqn{Xexpand}{
   X_n = \sum_{p=0}^\infty \lambda^{2p} X_{n,p} \,,
  }
 and now the $X_{n,p}$ are independent of $\lambda$.  Also $K$ itself
may be taken to have an expansion in $\lambda$,
  \eqn{Kexpand}{
   K = \sum_{q=0}^\infty \lambda^{2q} k_q \,.
  }
 Translating back into the notation of \ABCexpand, we have $k_0=k$,
$A_{0,0}=0$, $A_{0,1}=a_0$, $A_{1,0}=a_1$, $A_{2,0}=a_2$, and the same
for $B$ and $C$.  We do not see any reason for non-analytic terms in
$\lambda$ to appear at any order in \ABCagain\ or \Xexpand, or for odd
powers of $\lambda$ to appear in \Xexpand\ or \Kexpand.  If either
thing happened, then it would seem likely that there are additional
physical constants of integration to specify beyond $\eta$ and the
ratio of minimum and maximum radius for the horizon.\footnote{Provided
we take $C_{1,0}(1)=1$, it is still true that the $\lambda$ of this
section coincides with the one in \LambdaDef, to leading order in
$\lambda$ itself.  Working out a more precise translation is
straightforward once the $X_{n,p}$ and $k_p$ are known up to a given
order.}

Given the results of previous sections, it should be plausible that
the Einstein equations $R_{\mu\nu} = 0$ boil down to an infinite set
of ODE's of the form
  \eqn{XODEs}{
   {\cal L}^X_n [ X_{n,p} ] = S_{X_{n,p}} \,.
  }
 The left hand side represents a first or second linear differential
operator ${\cal L}^X_n$ operating on the $X_{n,p}$.  Linear equations
for $X_{n,p}$ arise at $O(\lambda^{n+2p})$.\footnote{Starting at
$O(\lambda^3)$, one also gets linear equations for functions $X_{n,p}$
which were determined at lower order.  These equations arise
multiplied by powers of $k_p$ and $x$: that is, they are artifacts of
expanding $\cos Kx$ and $\sin Kx$ in a basis of functions with period
$2\pi/k_0$.  It was certainly easy to check at $O(\lambda^3)$ that
these ``secular'' equations were automatically satisfied given the
$O(\lambda)$ equations; probably this could be demonstrated at all
orders, but we will not attempt complete rigor here.}  For $n\neq 0$,
the form of these differential operators is identical to the forms in
equations \bZeroSolve, \bpZeroSolve, \acOneODEs, only with $k$
replaced by $n k_0$.  The right hand side of \XODEs\ represents a sum
of terms, each expressible as a product of several $X_{n_i,p_i}$ and
their derivatives, as well as various $k_{q_j}$, subject to two sum
rules: $\sum_i s_i n_i = n$, where each $s_i = \pm 1$, and $\sum_i
(n_i+2p_i) + \sum_j q_j = n+2p$.  The first of these rules comes from
Fourier analysis, and the second comes from power counting in
$\lambda$.  This means that one may solve the equations \XODEs\ order
by order in $\lambda$ (that is, in order of increasing
$n+2p$)---modulo a difficulty to be discussed below.  At a given order
in $\lambda$, the equations with different $n$ are independent.  There
is a further useful property of the equations \XODEs, easily
demonstrated order-by-order: for $n \neq 0$, one can find algebraic
expressions for $B_{n,p}$ and $B_{n,p}'$ in terms of $A_{n,p}$,
$C_{n,p}$, and other $X_{n,p}$ of lower order, and first derivatives
of these quantities.  For $n=0$, this is not possible.

A subtler point is how we determine all the constants of integration,
including the $k_q$.  There is some freedom on how this is done,
corresponding to two obvious considerations: first, we have not
entirely fixed the diffeomorphism freedom with the ansatz \BSAnsatz\
and \ABCagain; second, starting at second order in $O(\lambda^2)$, one
has the freedom to simultaneously add something to the mass and change
$K$, in essence rescaling the solution by a perturbatively small
amount.  Even after demanding that $A(x,y)$, $B(x,y)$, and $C(x,y)$
vanish in the limit $y \to \infty$, there is some ambiguity left.  To
``fix a scheme,'' one could for example say $X_{0,0}=0$,
$C_{0,p}(1)=0$ for all $p$, $C_{1,0}(1)=1$, and $C_{1,p}(1)=0$ for
$p>0$.  Then the $k_q$ would be fixed as an integration constant in
the equations for $X_{1,q+1}$.  We will call this the standard scheme.

To make the discussion more definite, let us consider the standard
scheme through the first few orders in perturbation theory.  At
$O(\lambda)$, we have only the $X_{1,0}$ equations, and these fix two
integration constants, which in the standard scheme are $k_0$ and
$A_{1,0}(1)$.  At $O(\lambda^2)$, the requirement that $A(x,y)$ and
$C(x,y)$ should vanish as $y \to \infty$ fixes $A_{0,1}(1)$ and
$B_{0,1}(1)$, while $C_{0,1}(1)$ is set to zero as part of the
standard scheme.  This is an arbitrary resolution of the ambiguity we
encountered in section~\ref{OlambdaSquared}.  Also we have at
$O(\lambda^2)$ the equations for $X_{2,0}$, and asymptotic flatness
fixes the integration constants $A_{2,0}(1)$ and $C_{2,0}(1)$.
Finally, at $O(\lambda^3)$, we have the equations for $X_{1,1}$ and
$X_{3,0}$.  The latter are similar to the $X_{2,0}$ equations, and
asymptotic flatness fixes $A_{3,0}(1)$ and $C_{3,0}(1)$; the former
are similar to the $X_{1,0}$ equations, and in our scheme of setting
$C_{1,1}(1)=0$ they fix $k_1$ and $A_{1,1}(1)$.  As an illustration of
how schemes might be altered, we could leave $C_{0,1}(1)$ free in the
analysis of the $X_{0,1}$ equations, but set $k_1=0$ as well as
$C_{1,1}(1)=0$ in the analysis of the $X_{1,1}$ equations, which would
leave us with $C_{0,1}(1)$ and $A_{1,1}(1)$ as the constants of
integration which are fixed by asymptotic flatness in the analysis of
the $X_{1,1}$ equations.  This scheme is computationally
disadvantageous because when one changes $C_{0,1}(1)$, it is necessary
to re-integrate the $X_{0,1}$ equations before solving the $X_{1,1}$
equations.  Physically, the new scheme corresponds to holding fixed
the asymptotic size of the $S^1$ around which the black string is
wrapped, whereas the standard scheme allows this asymptotic size to
vary at $O(\lambda^2)$ but sets to zero a particular $O(\lambda^2)$
contribution to the change in mass.  Naively, one might propose yet
another scheme where the total mass is held fixed at $O(\lambda^2)$
and $k_1=0$ as well, but $C_{1,1}(1)$ is left free, in which case
$A_{1,1}(1)$ and $C_{1,1}(1)$ would be the constants of integration
fixed by analysis of the $X_{1,1}$ equations.  This seems natural, but
in fact we suspect that this scheme is degenerate, because it takes us
out toward non-zero $\lambda$ on a curve where $\eta$ is constant at
least to $O(\lambda^2)$.  The dependence of $\eta$ on $\lambda$ at
lowest non-trivial order is part of the physical ``output'' of our
numerics, and so should not be fixed as part of an arbitrary scheme.

Let us now to summarize our immediate aims without getting too tangled
up in schemes: we want to pin down the $O(\lambda^2)$ change in the
mass of the string which was left undetermined in
section~\ref{OlambdaSquared}.  To do it we have to proceed to
$O(\lambda^3)$.  A clever trick is to fix $(\delta e) / e$ arbitrarily
at $O(\lambda^2)$, but instead allow the size of the $S^1$ to change
at $O(\lambda^2)$.  That's equivalent because we can always rigidly
rescale the whole solution to bring the size of the $S^1$ back to what
it was, and in the process recover unambiguously the desired
$O(\lambda^2)$ contribution to $\delta e$.

So far, we have not discussed the $X_{2,0}$ equations in any detail;
indeed, the analysis of them is a rather uninteresting replay of the
$X_{1,0}$ equations.  Let us only remark that $B_{2,0}$ can be
determined algebraically, and the remaining two equations are second
order in $A_{2,0}$ and $C_{2,0}$, and involve source terms which can
be written as quadratic expressions in $A_{1,0}$, $C_{1,0}$, and their
first derivatives.  Regularity at the horizon fixes $A_{2,0}'(1)$ and
$C_{2,0}'(1)$ once $A_{2,0}(1)$ and $C_{2,0}(1)$ are known.  These
latter two quantities must be fixed arbitrarily to carry out a
numerical integration, and their values then are determined by the
requirement that both $A_{2,0}$ and $C_{2,0}$ shrink exponentially for
large $y$ rather than growing.  The result is
  \eqn{TwoNumericsResult}{
   A_{2,0}(1) = 0.34 \,, \qquad C_{2,0}(1) = -0.69 \,.
  }
 The $X_{2,0}$ equations are completely independent of the $X_{0,1}$
equations, so the arbitrariness of $C_{0,1}(1)$ does not affect
\TwoNumericsResult.

At $O(\lambda^3)$, the $X_{1,1}$ equations and the $X_{3,0}$ equations
are independent.  We have not analyzed the $X_{3,0}$ equations in any
detail, but clearly they will turn out just like the $X_{2,0}$
equations: $B_{3,0}$ will be determined algebraically, and
$A_{3,0}(1)$ and $C_{3,0}(1)$ will be fixed by normalizability at
infinity.  It is the $X_{1,1}$ equations that will primarily interest
us for the rest of this section.  Once again, $B_{1,1}$ can be
algebraically eliminated, and what is left is linear second order
equations for $A_{1,1}$ and $C_{1,1}$, sourced by expressions up to
cubic in the $X_{1,0}$, and also depending on $k_1$, $X_{0,1}$, and
$X_{2,0}$.  The form of the homogeneous equations (that is, with the
source terms set artificially to zero) is {\it identical} to
\acZeroODEs, only with $a_1$ replaced by $A_{1,1}$ and $c_1$ by
$C_{1,1}$ (this is a general fact, indicated in \XODEs\ by the
dependence of ${\cal L}^X_n$ only on $X$ and $n$, not $p$).  In our
standard scheme, $k_1$ and $A_{1,1}(1)$ are determined by
normalizability of $A_{1,1}$ and $C_{1,1}$ at infinity, so there is a
two-parameter shooting problem.  The results of numerics are
  \eqn{ThirdResult}{
   k_1 = 0.70 \,, \qquad A_{1,1}(1) = -0.24 \,.
  }

The best way to extract the effect on the mass is to return to our
definition of $\eta$.  For the five-dimensional black string, the
``average'' Schwarzschild radius appearing in \EtaDef\ is proportional
to the average energy density $e$, so $\eta$ can be expressed as $eK$
up to a factor which includes a power of Newton's constant.  Combining
the $O(\lambda^2)$ effects from \DeltaEDeltaS\ and \ThirdResult\ (both
of which were computed in the standard scheme), we find
  \eqn{etaChange}{
   {\delta \eta \over \eta} = {\delta e \over e} + {\delta K \over K}
     = \eta_1 \lambda^2 \qquad\hbox{where}\quad \eta_1 \approx 1.45 \,.
  }
 Because $\eta$ is independent of rigid rescalings of the entire
solution, $\eta_1$ should be scheme-independent.  In fact, there is
approximately $0.6\%$ variation in $\eta_1$ when $C_{0,1}(1)$ is
changed from $0$ to $1$, and much smaller variation when $C_{1,1}(1)$
is changed from $0$ to $1$.

It is possible to extend the First Law argument of
section~\ref{OlambdaSquared} to a computation of the entropy
difference of the uniform and non-uniform solutions to $O(\lambda^4)$.
Again holding the asymptotic size of the $S^1$ fixed, let us integrate
the infinitesimal form of the First Law along the curve of non-uniform
black brane solutions, starting from the point where it joins onto the
uniform solutions, and assuming expansions
  \eqn{TSExpand}{
   M = \sum_{p=0}^\infty M_p \lambda^{2p} \qquad
   S = \sum_{p=0}^\infty S_p \lambda^{2p} \qquad
   T = \sum_{p=0}^\infty T_p \lambda^{2p} \,.
  }
 Starting with $M_0 = 2 T_0 S_0$ (true because at $\lambda=0$ the
equation of state is the same as for the four-dimensional
Schwarzschild black hole) one quickly obtains $M_1/M_0 = S_1/(2S_0)$
and $M_2/M_0 = S_2/(2S_0) + T_1 S_1/(4T_0 S_0)$.  Now let us inquire
how the entropy of a {\it uniform} string changes if we change the
mass by $\Delta M$.  Since $S \propto M^2$, we have $(\Delta S)/S_0 =
2(\Delta M)/M_0 + [(\Delta M)/M_0]^2$.  What we really want is the
difference between the entropy of a non-uniform string and a uniform
one of the same mass.  So we set $\Delta M = M_1 \lambda^2 + M_2
\lambda^4$ and obtain, to $O(\lambda^4)$,
  \eqn{SUNdifference}{\eqalign{
   &{S_{\rm non-uniform} - S_{\rm uniform} \over S_{\rm uniform}} = 
    {S_{\rm non-uniform} - S_{\rm uniform} \over S_0}  \cr
   &\qquad\qquad{}= {S_1 \over S_0} \lambda^2 + {S_2 \over S_0} \lambda^4 - 
     {2\Delta M \over M_0} - \left( {\Delta M \over M_0} \right)^2  
     = -\left[ {T_1 M_1 \over T_0 M_0} + 
          \left( {M_1 \over M_0} \right)^2 \right]
      \lambda^4 \cr
   &\qquad\qquad{}= -\left( {\delta T \over T} - 
      {\delta K \over K} \right) {\delta\eta \over \eta} - 
     \left( {\delta\eta \over \eta} \right)^2 \,.
  }}
 The first equality holds, even though $S_{\rm uniform}$ deviates from
$S_0$ by $O(\lambda^2)$, because the numerator is an $O(\lambda^4)$
quantity.  In the final equality, we have used the fact that $\eta$ is
proportional to $M$, and the constant of proportionality is fixed if
the asymptotic size of $S^1$ is held fixed.  We have also used the
facts that $\lambda^2 T_1/T_0 = (\delta T)/T$ when the
asymptotic size of $S^1$ is held fixed, and that $(\delta T)/T -
(\delta K)/K$ is a scheme-independent quantity at $O(\lambda^2)$.  The
final two expressions in \SUNdifference\ involve only quantities
computed at $O(\lambda^2)$, so with \ThirdResult\ in hand one can
compute, to the relevant orders,
  \eqn{SUNresult}{\seqalign{\span\TC}{
   {\delta T \over T} - {\delta K \over K} \approx -1.04 \lambda^2 \cr
   {S_{\rm non-uniform} - S_{\rm uniform} \over S_{\rm uniform}} =
    \sigma_2 \lambda^4 \qquad\hbox{where}\quad \sigma_2 \approx
    -0.59 \,.
  }}
 The coefficients quoted were computed in the standard scheme.  There
is approximately $3\%$ variation in $\sigma_2$ when $C_{0,1}(1)$ is
changed from $0$ to $1$, and much smaller variation when $C_{1,1}(1)$
is changed from $0$ to $1$.  The value of $\sigma_2$ could be used as
a consistency check on numerics at $O(\lambda^4)$, just as the
near-vanishing of $\sigma_1$ was used in section~\ref{OlambdaSquared}
as a consistency check at $O(\lambda^2)$.

The $0.6\%$ variation in $\eta_1$ upon changing $C_{0,1}(1)$ from $0$
to $1$, and the $3\%$ variation in $\sigma_2$, are indications of the
size of numerical errors, since in principle the quantities in
\SUNresult\ are scheme-independent.  Additively, the variation in
$\sigma_2$ is only twice as large as the variation in $\eta_1$, and it
is nearly the same as the variation in $\sigma_1$ discussed in the
paragraph following equation \DeltaES.  The upshot is that the
variations observed in these three quantities are consistent with one
another, and small enough that we can be quite confident that $\eta_1
> 0$ and $\sigma_2 < 0$.

\subsection{Improving and extending the numerics}
\label{Improve}

There are various ways in which the numerics discussed in previous
sections could be improved and extended.  To begin with a rather
technical point, we have run all our numerics with Mathematica's
built-in {\tt NDSolve} routine, which uses adaptive step-sizing: the
discrete steps can change each time one solves a differential equation
with different initial conditions.  This results in decreased
stability for shooting algorithms which became particularly noticeable
in the analysis of the $X_{2,0}$ equations.  Fixing the step size once
and for all would have its advantages where stability is concerned,
although one would probably want to retain the feature that shows up
with {\tt NDSolve}, that the steps are smaller near the horizon.
Another technical point which could clearly stand some improvement is
the treatment of initial conditions.  The rough-and-ready method we
have used is to observe that second derivative terms cancel out of the
differential right at the horizon, so evaluating the equations there
results in a constraint on first derivatives---but one which we then
impose at a slight distance away from the horizon (on the order of
$10^{-4}$) so that {\tt NDSolve} won't be faced with excessively small
coefficients on the second derivative terms. A better method, though
more labor-intensive, would be to obtain analytic approximations to
the regular solutions to some moderate order, and use them as the seed
for numerics.  (This is straightforward in principle because the
horizon is a regular singular point of the differential equations).
It might even be hoped that if one knew the near-horizon asymptotics
as well as the long-distance tails to sufficiently high order, a
reasonably uniform approximation to the $X_{n,p}$ could be obtained
via matching.  This is not an approximation that could be controlled
by a small parameter, as it is in black hole absorption at low energy.
It might nevertheless be useful for large $n$, where numerics gets
harder and harder due to the increasingly powerful exponentials in the
growing solutions.

To determine $C_\infty$ at any even order in $\lambda$, we have to
distinguish a $(\log y)/y$ behavior from a $1/y$ behavior in a
function, $c_0(y)$, known only numerically.  This is one of the
hardest aspects of the numerics, so it is natural to suspect that it
is a dominant source of numerical error.  The quantities $\sigma_1$,
$\eta_1$, and $\sigma_2$ all depend on evaluations of $C_\infty$, and
all three displayed variations of similar magnitude (of order
$10^{-2}$) when the scheme was changed.  On the other hand, spot
checks of the values of the components of the Ricci tensor gave values
on the order of $10^{-15}$ up to $O(\lambda^3)$, when numerical
evaluation was performed after elimination of all second derivatives.
This is only ten times the working precision for most executions of
{\tt NDSolve}.  These numerical observations are consistent with the
hypothesis that the determination of $C_\infty$ dominates the
numerical error.\footnote{In the original hep-th version of this
paper, a different source of numerical error dominated: the numerics
on the second and third equations of \acZeroODEs\ was unstable near
the horizon, leading to a poor value for $a_0(1)$.  This is an
instance where matching an analytic near-horizon solution to a
numerical one would have avoided any problem.}

In the end, customized code would be needed to optimize results.  We
have aimed in this paper to avoid excessively hard-core numerics,
extracting instead what information we could from standard tools.  The
total amount of CPU time needed to check all the numbers presented in
this paper is roughly an hour on a 800 MHz Pentium PC.

\subsection{Interpreting the numerics}
\label{Interpret}

Broadly speaking, the goal of this paper has been to make some dent in
the problem of constructing stationary non-uniform black brane
solutions and elucidating their thermodynamic properties.  Our point
of entry was the observation that there is a zero-mode fluctuation for
the uncharged black string in five dimensions, with wave-number $k =
0.876/R_{\rm Schwarzschild}$.  This suggests that there is a
one-parameter family of non-uniform solutions in the $\eta$-$\lambda$
plane that joins onto the uniform solutions at a ``critical point,''
$(\eta_c,\lambda_c) = (0.876,0)$.  We may then ask, 
  \begin{enumerate}
   \item What curve in the
$\eta$-$\lambda$ plane is traced out by these non-uniform solutions?

   \item What is the entropy along this curve?

   \item How does a black string behave in real time as one slowly
adds mass or allows mass to slowly Hawking radiate away, such that
$\eta$ passes through $\eta_c$?
  \end{enumerate}
 A perturbative expansion in small $\lambda$ seems well-suited to
study the first two questions near the critical point.  Given the
answers to these two questions, we can apply the Second Law plus some
``natural'' assumptions to give a qualitative answer to the third.

\begin{figure}
   \centerline{\psfig{figure=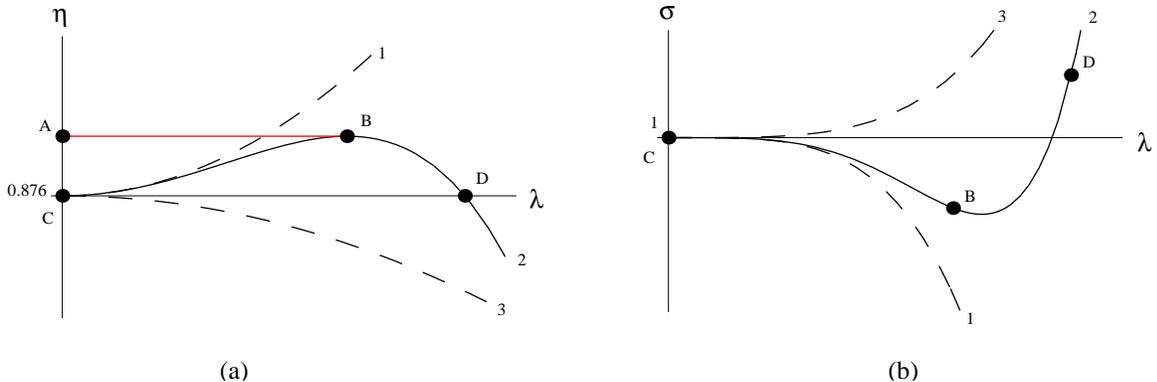,width=6in}}
   \caption{(a) Possible curves, drawn schematically in the
$\eta$-$\lambda$ plane, describing stationary, uncharged, non-uniform
black strings in five dimensions.  The solid curve is the likeliest.
(b) $\sigma$ versus $\lambda$ for these non-uniform strings.  Again,
the solid curve is the likeliest.}\label{figB}
  \end{figure}

Possible answers we could envisage to questions~1 and~2 are sketched
in figure~\ref{figB}.  Given the analyticity properties in $\lambda$
of the perturbation expansion, we expect
  \eqn{EtaSigmaExpand}{\seqalign{\span\TC}{
   \eta = \eta_c + \sum_{p=1}^\infty \lambda^{2p} \eta_p  \cr
   \sigma \equiv {S_{\rm non-uniform} \over S_{\rm uniform}} = 
    1 + \sum_{p=1}^\infty \lambda^{2p} \sigma_p \,,
  }}
 for some coefficients $\eta_p$ and $\sigma_p$.  In the first line of
\EtaSigmaExpand, as in \SUNdifference\ and \SUNresult, $S_{\rm
non-uniform}$ and $S_{\rm uniform}$ are the entropies of black string
solutions with the {\it same mass}.\footnote{Again we are glossing
over a difference in the simplest conceptual definition of $\lambda$,
namely \LambdaDef, and the computationally convenient definition of
$\lambda$, which in the standard scheme could be expressed as the
Fourier coefficient of $\cos kx$ in the fractional deviation of the
local Schwarzschild radius from its mean value.  We do not think this
ambiguity will affect the issue of analyticity, but clearly it would
change the values of the coefficients in \EtaSigmaExpand.}
 The First Law guarantees that $\sigma_1 = 0$, and this was verified
to good accuracy by the numerics through $O(\lambda^2)$ (see \DeltaES\
and the discussion below it).  Numerics through $O(\lambda^3)$ led to
a positive result for $\eta_1$ (see \etaChange).  Combining these
numerics with First Law considerations further predicts $\sigma_2 < 0$
(see~\SUNresult).  Thus by our calculations of the previous few
sections, we rule out the curves marked (3) in figure~\ref{figB}.

Although we have no direct evidence, it would seem sensible that $\eta
< \eta_c$ and $\sigma > 1$ for sufficiently large $\lambda$.  The
first inequality guarantees that, for some range of $\eta$ less than
$\eta_c$, there are stationary solutions with finite $\lambda$ and a
single maximum ($n=1$).  The second inequality says that these
solutions are accessible as the endpoint of Gregory-Laflamme evolution
for an unstable uniform string of nearly the same mass.  Taking these
inequalities as working assumptions, we rule out the curves marked (1)
in figure~\ref{figB}(a) and~\ref{figB}(b), and settle on the curves
marked (2) as the likeliest.\footnote{Of course, one might also
imagine the more arcane possibility that curve (1) is right and that
there is a new disconnected branch of non-uniform solutions at finite
$\lambda$ which extends below $\eta_c$.}  We have incorporated an
additional prejudice into curve (2) of figure~\ref{figB}(b): namely,
we expect that the non-uniform black string with maximal $\eta$ should
have an entropy less than a uniform black string of the same mass, so
that if one started with such a non-uniform solution and added mass,
it would have someplace to go (namely, to a uniform solution).  We
emphasize that our current numerics determines {\it only} the leading
non-zero behavior of $\eta$ and $\sigma$ for small $\lambda$.  The
rest of our expectations are motivated by the hope that once we know
the continuous moduli space of static non-uniform solutions, we will
be able to give a complete qualitative account of the real-time
dynamics in the vicinity of $\eta_c$.

If indeed the curves marked (2) are correct, then the transition
between uniform and non-uniform strings is first order.  (The
transition is also first order if there is a discontinuous moduli
space, as discussed in the previous footnote.  We will not consider
this option further since we have no way of exploring it).  Suppose we
start with a very massive uniform black string wrapped on an $S^1$
whose size we hold fixed.  Then as the string Hawking radiates, $\eta$
decreases from its initial large value.  When it reaches $\eta_A$,
there is another solution available, but provided fluctuations are
small and tunneling is suppressed ({\it i.e.}{} $R_{\rm Schwarzschild}
\gg \ell_{\rm Pl}$) the system will be unable to jump from (A) to (B)
even if (B) has higher entropy (which it well may not, since $\sigma_2
< 0$).  Instead it will proceed down to (C), become locally unstable,
and evolve in some non-adiabatic fashion, settling down presumably to
(D), which is possible since we have assumed in figure~\ref{figB}(b)
that (D) has larger entropy than (C).\footnote{Since there is some
non-adiabatic behavior, a finite amount of mass will be lost to {\it
classical} gravitational radiation.  However, this lost mass will
scale as some positive power of the characteristic frequency of the
transition.  By scaling up the size of the whole system, we can make
this loss controllably small.}  Going the other way, if one starts
with a non-uniform string with $\eta < \eta_c$ and slowly increases
its mass (for instance by feeding it dust), then the system would
reach point (B) and then evolve non-adiabatically to (A).  Again this
is possible since we have assumed in figure~\ref{figB}(b) that the
entropy at (B) is less than at (A).  We are tempted to say that the
latent heat of the transition from uniform to non-uniform solutions
and back would be determined by $\sigma$ at points (B) and (D), but
this seems a slight misnomer since in the usual language of phase
transitions, both jumps (from (C) to (D) and from (B) to (A)) would be
best described as discontinuous transitions from a meta-stable phase
to a stable one.

To test the picture suggested in the previous paragraph, it would be
very interesting to go to $O(\lambda^5)$ in the numerics and obtain
the constants $\eta_2$ and $\sigma_3$.  A certain range of negative
values of $\eta_2$ and positive values of $\sigma_3$ would support the
hope that $\sigma < 1$ at (B) at $\sigma > 1$ at (D).  Unfortunately,
because (B) and (D) are both at finite $\lambda$, there is a
possibility that low order perturbative results will mislead.  Also,
one would have to be careful to use an unambiguous definition in
$\lambda$.  In computing lowest order results, it's OK that we used
interchangeably two different definitions of $\lambda$ which differ
multiplicatively by $1+O(\lambda^2)$.  That won't work at higher
orders.

One might imagine situations in which the curves marked (3) in
figure~\ref{figB}(b) are qualitatively correct---for instance, in a
different dimension, or with some form of charge on the string.  Then
the transition is probably continuous: a uniform black string crossing
through $\eta_c$ will smoothly develop non-uniformity by moving onto
the branch of moduli space where $\lambda\neq 0$.  The Second Law
permits this provided $\sigma > 1$, as we assumed in curve (3) of
figure~\ref{figB}(b).  If a continuous transition were found for some
black brane system, it would be interesting to compute critical
exponents, though naively it seems pretty likely that everything of
interest will be analytic in $\lambda$, perhaps even $\lambda^2$.

Another way to describe the difference between a first order and
continuous transition, as we are using the terms, is that in a
continuous transition, the amount of ``thrashing,'' or non-adiabatic
evolution, that an initially uniform string below $\eta_c$ experiences
in its real time evolution before settling down to a static
non-uniform solution, would vanish as the initial $\eta$ approaches
$\eta_c$; whereas in a first order transition there is a finite
minimum amount of thrashing.

Stranger possibilities than curves (2) or (3) may be contemplated: for
instance, if the curves marked (1) are correct, then the non-uniform
solutions in the moduli space we have explored are never reached
starting from a Gregory-Laflamme instability, both because they have
the wrong mass and because their entropy is too low.  This scenario is
not ruled out by our numerical results, but we would regard it as very
surprising.  It's worth noting that real-time simulations of
gravitational collapse could lead to static non-uniform solutions {\it
regardless} of which curves are right in figure~\ref{figB}.

Clearly we are in the realm of the speculative, and would be
considerably aided by results at higher order in $\lambda$.  Some form
of real time numerics would also be desirable in order to get to
finite $\lambda$.  An interesting point which might be cleared up by
analytical means is whether one can show that non-uniform strings do
not exist above a certain $\eta$.  It seems likely that this is true,
since for very large $\eta$ we could perform a Kaluza-Klein reduction
along the $S^1$ and then try to argue that static spherically
symmetric black holes in the resulting four-dimensional theory are
unique.\footnote{We thank G.~Horowitz for raising this point.  Note,
this line of argument might be more subtle than it naively appears,
since if static non-uniform solutions exist at all in five dimensions,
the four-dimensional uniqueness argument cannot be true for all
$\eta$, but must somehow hinge on the relative size of typical
Kaluza-Klein masses and the inverse Schwarzschild radius.}

\section{Other black branes}
\label{Discuss}

Although relatively unexplored, the phenomena associated with
non-uniform black branes promise to be quite interesting and varied.
The main difficulty is the dearth of problems that can be attacked
without the help of heavy-duty numerics.  Thus, at this stage, we have
many more questions than answers.

The main question we have addressed in this paper is the nature of the
transition from uniform to non-uniform solutions at the critical size
where the Gregory-Laflamme instability first shows up.  Focusing on
the uncharged black string in five dimensions, we have argued that
although there is a continuous moduli space of solutions, with a
branch of non-uniform solutions merging onto the uniform ones at
$\eta=\eta_c$, the shape of the curve in the $\eta$-$\lambda$ plane is
probably such that there is a first order transition.

Since our calculation was clearly quite specific to black strings in
five dimensions, let us now consider what our expectations might be
for more general black branes, based on symmetries and genericity.
Before we begin, it is worth considering the analogy with the Landau
treatment.\footnote{This analogy was developed in part through
conversations with participants of ``Avatars of M-theory'' and
subsequently with E.~Witten.}  Generally speaking, crystallization is
a first order transition, and the reason is that there are cubic terms
in the Landau free energy.  More explicitly (see for example \cite{CL}
for even more detail) one imagines starting with a uniform phase of
some substance which at some temperature manifests an instability in
density perturbations at a particular wavelength, call it $k$.  We
might write the density perturbations in terms of a scalar field,
$\phi \sim \delta\rho$.  Then the instability is toward developing
some VEV for $\phi = \phi_{\vec{k}} \cos (\vec{k} \cdot \vec{x})$.  We
might suppose that the instability in $\phi_{\vec{k}}$ develops in a
smooth fashion---for instance, via terms in the Landau free energy of
the form
  \eqn{LandauQuadratic}{
   {\cal F}_{2+4} = \sum_{|\vec{k}|=k} \left( r(T) \phi_{\vec{k}} 
    \phi_{-\vec{k}} + u (\phi_{\vec{k}} \phi_{-\vec{k}})^2 \right) \,,
  }
 where $u>0$ and $r(T)$ passes through zero at the crystallization
temperature.  If \LandauQuadratic\ were all, then the transition would
be second order.  But translation invariance also allows cubic terms
of the form
  \eqn{LandauCubic}{
   {\cal F}_3 = \sum_{\vec{k}_1 + \vec{k}_2 + \vec{k}_3 = 0 \atop
    |\vec{k}_i| = k} g \phi_{\vec{k}_1} \phi_{\vec{k}_2} 
     \phi_{\vec{k}_3} \,.
  }
 Such terms are allowed provided there is no $\phi \to -\phi$
symmetry, and provided that we can construct equilateral triangles of
allowed momenta, $\vec{k}_1$, $\vec{k}_2$, and $\vec{k}_3$.  If the
cubic terms \LandauCubic\ are present, then inevitably there is a
first order phase transition to a state where several different
wave-numbers condense simultaneously, with phases set in such a way as
to make ${\cal F}_3 < 0$.  We will refer to this as ``phase locking.''
In fact, one might further expect that the preferred crystal structure
in dimensions higher than two (where there is only one natural
candidate emerging from \LandauCubic, namely the hexagonal lattice)
will be the one with the most equilateral triangles per unit cell.
This actually tallies with reality in three dimensions for crystal
structures near their melting point: a great many are BCC.

The non-uniform black hole problem is similar to the model of
crystallization described above in two important respects: 
  \begin{itemize}
   \item[1.] Non-zero wavelength modes are unstable.
   \item[2.] There is nothing like a $\phi \to -\phi$ symmetry, since
it's obviously different to make the black hole horizon bigger than to
make it smaller.
  \end{itemize}
 But it's different in three important respects:
  \begin{itemize}
   \item[3.] {\it All} wavenumbers below a critical one are unstable.
   \item[4.] We must work in the microcanonical ensemble.
   \item[5.] For large, smooth, classical solutions, no tunneling is
allowed.
  \end{itemize}
 Compactifying the dimensions parallel to the black hole horizon is a
convenient way of controlling point 3): we can contrive to deal with
only one unstable mode at first (just below $\eta_c$), then try to
work up to several unstable modes at smaller $\eta$.  Compactification
has the additional advantage that one ducks out of any
Coleman-Mermin-Wagner worries about whether long-range order is
possible in low dimensions.

Based on the computations in section~\ref{Numerics}, it appears that
the black string transition is first order, but for reasons having
nothing to do with cubic terms like \LandauCubic, which are of course
impossible in one spatial dimension.  A closer analogy in a Landau
treatment would be a free energy like ${\cal F} = r(T) \phi^2 + u
\phi^4 + v \phi^6$, where $u$ is {\it negative}, $v$ is positive, and
$r(T)$ varies.  For such a free energy, if tunneling and bubble
nucleation is suppressed, a system that starts at $\phi=0$ for
positive $r(T)$ stays there until $r(T)=0$, then rolls off to a lower
minimum at finite $\phi$.  For black strings in higher dimensions, or
for charged black strings, we would not be at all surprised to find
second order transitions.  The positivity of $\eta_1$, like the
negativity of $u$ in the analogy above, seems more likely to be a fact
specific to the five-dimensional black string than a general property
of all black branes experiencing a Gregory-Laflamme instability.

For $p$-branes with $p>1$, can we infer that transitions from uniform
to non-uniform branes should be generically first order?  The answer
depends on what one really means by a first order transition for black
branes.  What does seem likely is that the moduli space of non-uniform
branes for $p>1$ will extend both above and below $\eta=\eta_c$, due
to phase-locking effects.  One could then imagine tunneling from
uniform to non-uniform branes at some $\eta > \eta_c$---that is,
before the uniform branes become unstable.  This would be the closest
analogy to what is usually meant by a first order transition in
condensed matter physics.  However, if we restrict ourselves to
systems which evolve classically and have only small fluctuations,
then tunneling is suppressed, and one can only ask what happens to a
uniform brane once it starts experiencing the Gregory-Laflamme
instability.  As discussed above, our preferred notion of a first
order transition in this setting is a transition that takes place
non-adiabatically and results in a finite minimum non-uniformity.
Then what determines whether a transition is first order or continuous
is whether $\eta > \eta_c$ or $\eta < \eta_c$ for small non-zero
$\lambda$.  In a perturbative treatment, we expect that the leading
contribution to $\eta$ will generically be at $O(\lambda^2)$.  For
$p>1$, it seems likely that there are odd powers of $\lambda$
contributing to $\eta$; however we cannot see how an $O(\lambda)$
contribution would arise: such contributions, it seems, could be
soaked up into a redefinition of $\eta_c$.  So a determination of the
$O(\lambda^2)$ contribution to $\eta$, similar to the computation of
$\eta_1$ in \etaChange, is the crucial indicator of first or second
order behavior.  If this contribution is positive, then a curve like
(2) of figure~\ref{figB} would pertain, and the transition is first
order; if negative, then a curve like (3) would pertain, at least for
small $\lambda$, and the transition is continuous.

Of course, if one is in a regime where tunneling is appreciable, then
what's most relevant in all cases (including the black string) is
whether there are, at a given value of $\eta$, more entropic solutions
at any finite $\lambda$ than the uniform black string.  But in such a
regime, other effects, like Hawking radiation and the classical
gravitational radiation during non-adiabatic evolution, also become
appreciable.

This paper has been focused largely on $\eta$ close to $\eta_c$---that
is, on black strings compactified on circles on the order of their
Schwarzschild radius.  Let us now turn our attention to small $\eta$,
{\it i.e.}{} the limit where the black string or brane is
decompactified.  Then there would be many unstable modes on a uniform
black string or brane (infinitely many in the strict $\eta \to 0$
limit).  But, as is evident from Figure~1 of \cite{glOne}, the {\it
most} unstable mode (that is, the one with maximum $\Omega$) is at
approximately $k=k_c/2$, where $k_c$ corresponds to the shortest
wavelength that is unstable.  The dispersion relation for the unstable
modes, as computed numerically in \cite{glOne}, is (very
approximately) $\Omega \approx {k_c \over 10} \sin \pi k/k_c$.  Thus
if one starts with a uniform black string with very small $\eta$, the
``dominant'' instability has finite wavelength, and one might
optimistically expect a crystal structure to form with lattice spacing
$a \approx 4\pi/k_c$.  A serious issue, however, is whether any
crystal structure would be stable.  Intuitively speaking, there are
two potential threats to stability:
  \begin{enumerate}
   \item Long-wavelength phonons: at wavelengths much long than
$4\pi/k_c$, it might seem plausible that something like the
Gregory-Laflamme instability still operates, tending to cluster more
mass in one region (populated by many maxima) at the expense of
another.  Actually, such an effect might even manifest itself at $\eta
\lsim \eta_c/2$, where we could start with a black string (say) with
two identical maxima and see whether the system is stable toward one
maximum growing while the other shrinks.

   \item Short-wavelength instabilities: if the ``necks'' between
maxima get too long and slender, then the local scale of
Gregory-Laflamme instabilities might be pushed small enough to occur
locally in the neck.  This might for instance create new miniature
local maxima in the nooks between the big ones.
  \end{enumerate}
 By focusing on solutions with $\eta$ close to $\eta_c$, we have
avoided the long-wavelength phonon problem entirely: they are
projected out by finite volume.  As long as $\lambda$ is less than
about $1$, we can also be fairly confident that short-wave
instabilities don't crop up.  However, if there really is first order
behavior in the five-dimensional black string, rather than a
continuous transition at $\eta = \eta_c$, then we cannot completely
rule out the possibility that short-wavelength effects will destroy
the stability of the $n=1$ solutions.  The criterion for these UV
instabilities to show up is a region of the black brane which is
considerably less ``thick'' (as measured by its average Schwarzschild
radius) than it is wide (as measured in the directions parallel to the
horizon).

If the instabilities described above are absent and there is a
genuinely stable crystal structure at $\eta = 0$, then of course the
physics at finite $\eta$ would be largely determined by frustration:
if a black brane is compactified on a torus which is commensurate with
its stable crystal structure, then the torus is just filled with a
region of undeformed crystal; whereas for other sizes or shapes, the
system fits in about as many unit cells as it can, with some strain
because of the boundary conditions, and generically first order
transitions between $n$ maxima and $n+1$.  The simplest reason that
such transitions should be first order is that they have different
symmetry groups, neither of which is a subgroup of the other.  For
instance, in the case of a black string, the symmetry group of a
solution with $n$ identical maxima is a semi-product of the cyclic
symmetry group ${\bf Z}_n$ and the reflection symmetry group ${\bf
Z}_2$.

At this point, we do not see a truly compelling reason to believe that
the instabilities described above are absent at small $\eta$ for the
generic non-uniform black brane solution.  If they are present, then
there would be no foreseeable endpoint to the evolution of an unstable
black brane that extends over a volume much greater than its
Schwarzschild radius.  The evolution might, for instance, pass
``fairly close'' to a crystal structure, but then be driven away by a
long-wavelength phonon instability.  It would be difficult to check
whether this happens via straightforward numerical solution of
Einstein's equations in real time.  The reason is that one needs an
arbitrarily large range of the coordinate we have called $x$ in order
to include the effects of soft phonons.  If numerics show that stable
solutions form at $\eta = 0.9 \eta_c$, and at $\eta = 0.5 \eta_c$, and
at $\eta = 0.1 \eta_c$, a skeptic might still point out that at $\eta
= 0.01 \eta_c$ there are ten times as many phonon modes available, and
the softest could be tachyonic.  Perhaps one could work at a series of
finite but small values of $\eta$ and then try to extrapolate the
observed phonon dispersion relation $\omega(k)$ down to zero
wavenumber.  The short-wavelength instabilities seem somewhat less of
a worry: if they were shown to be absent for finite $\eta$ (say in a
real-time numerical treatment of $n=1$ solutions) then it would be a
surprise to find them at small $\eta$ unless triggered somehow by a
soft phonon instability.

To improve the analogy with crystallization, one could consider a
Landau free energy in which all sufficiently soft modes of $\phi$ are
unstable at quadratic order.  What structures form and are stable is
then highly dependent on the structure of the $\phi^3$ and $\phi^4$
terms.  One could contrive, for instance, for a VEV of the most
quadratically unstable mode to stabilize all the other modes---or for
such a VEV to actually destabilize modes which were stable at
$\phi=0$.  The possibilities for non-uniform black branes seem, at
this point, potentially as varied and interesting.  It would be a
fascinating enterprise (though, most likely, a computationally
intensive one) to investigate which of the possibilities are realized
by simple gravitational lagrangians.

Some final remarks are in order regarding the case of charged black
branes.  For a black $p$-brane, we wish to consider only charge under
a $p+1$-form gauge potential.  The charge determines a definite length
scale $R_{\rm charge}$ to which which the horizon radius $R_{\rm
horizon}$ should be compared.  If $R_{\rm horizon} \gg R_{\rm
charge}$, then the physics should be essentially the same as for the
uncharged case.  When $R_{\rm horizon} \sim R_{\rm charge}$, the black
brane often becomes locally thermodynamically stable.  It is has been
argued in \cite{gMitraOne,gMitraTwo} that for locally
thermodynamically stable branes, there is no perturbative
Gregory-Laflamme instability, no matter how small $\eta$ may be; and
it was further conjectured that the Gregory-Laflamme instability
arises precisely when local thermodynamic stability is lost.  Some
evidence for this claim was presented in \cite{gMitraOne,gMitraTwo},
and further arguments have appeared in \cite{Reall,Rangamani}.  A
close look at these papers suggests that when the instability first
arises (that is, when the black branes are just massive enough to be
unstable), the wavelength of the unstable modes is extremely large.
The short-wavelength instabilities described above are less of a
worry: regions where the mass is decreased will eventually fall into
thermodynamic stability, and then it can't have any local
instabilities at all.

It is perhaps worth reviewing the argument \cite{gMitraOne,gMitraTwo}
for why local thermodynamic stability precludes a perturbative
Gregory-Laflamme instability.  We emphasize ``local'' and
``perturbative'' here because it is always possible to increase the
entropy of an infinitely extended, near-extremal charged brane by
moving all the non-extremal mass into enormous, sparsely distributed
Schwarzschild black holes through which the otherwise extremal charged
brane now passes.  The essence of local thermodynamic stability (which
for a $p$-brane with no other quantum numbers than its mass and charge
is simply the positivity of the specific heat) is that if one keeps
average mass density constant but tries to make the system slightly
non-uniform, the entropy goes down.  Thus from the often-invoked Second
Law, we learn that a Gregory-Laflamme instability is not possible.
The point of the conjecture \cite{gMitraOne,gMitraTwo} that a
Gregory-Laflamme instability arises as soon as local thermodynamic
stability is lost is that an infrared instability would seem plausibly
to be entirely driven by thermodynamic considerations.

\section{Conclusions}
\label{Conclude}

Although there seems to be a continuous moduli space of non-uniform
black branes connecting to the uniform ones at the critical mass
density where a Gregory-Laflamme density sets in, we have seen that
this is not enough by itself to guarantee a continuous phase
transition from uniform to non-uniform solutions.  The results of our
numerical study essentially boil down to two numbers, $\eta_1$ and
$\sigma_2$.  The positivity of $\eta_1$ (see \etaChange) means that
slightly non-uniform solutions are {\it above} the critical mass
density, by an amount proportional to the square of the amplitude of
the horizon fluctuations, which we have called $\lambda$.  The
negativity of $\sigma_2$ (see \SUNresult) means that these non-uniform
solutions have {\it lower} entropy than uniform black strings of the
same mass, by an amount proportional to the fourth power of $\lambda$.
These results are exactly contrary to our original expectations, which
were based on the hope that there would be a continuous transition
describable as motion along the moduli space.  The simplest scenario
consistent with our numerical results, together with the general
expectation \cite{hm} that the evolution of unstable branes settles
down to a static endpoint, is that the moduli space of non-uniform
solutions eventually curves down in the $\eta$-$\lambda$ plane so as
to provide the desired static endpoint, but at finite $\lambda$; and
that for the non-uniform solutions which are below the critical mass
density, the entropy is higher than that of a uniform string of the
same mass (see figure~\ref{figB}).  If all this is true, then as soon
as the mass of a five-dimensional, uncharged black string falls below
a critical threshold, the Gregory-Laflamme instability leads to
non-adiabatic evolution to a finitely non-uniform solution.  This is
what we regard as a first order transition.

How trustworthy are the numerics?  Clearly they are not perfect, since
$\eta_1$ and $\sigma_2$ fluctuate, respectively, by $0.6\%$ and $3\%$
as we ``change the scheme'' by an $O(1)$ quantity (as described more
precisely in section~\ref{HigherOrder}).  Abstractly, changing scheme
means that we make some combination of a rescaling of the solution and
a diffeomorphism on the $x$ and $y$ coordinates.  The quantities
$\eta_1$ and $\sigma_2$ are supposed to be scaling- and
diffeomorphism-independent.  It is quite plausible that the
percent-level fluctuations we observe in $\eta_1$ and $\sigma_2$ are
due to numerical error, particularly in the evaluation of the quantity
we called $C_\infty$.  We can say with some confidence that we got the
signs right on $\eta_1$ and $\sigma_2$.  Since the claim of a first
order rather than continuous transition for the five-dimensional black
string is dependent only on the signs, we believe this is a robust
result.

Extending the numerics to higher orders is a project that we hope to
report on in the future.  We also hope that it may become possible to
address some of the issues of stability raised in
section~\ref{Discuss} at a sufficiently high order in perturbation
theory.

\section*{Acknowledgments}

We thank J.~Schwarz, N.~Warner, E.~Witten, and especially G.~Horowitz
for useful discussions.  We are particularly grateful to T.~Wiseman
for pointing out and correcting two errors in the original version of
the manuscript, leading to considerably improved numerical accuracy.
This work was supported in part by the DOE under grant
DE-FG03-92ER40701.

\bibliography{bead}
\bibliographystyle{ssg}

\end{document}